\documentclass[aps,prd, twoside,superscriptaddress,showpacs, preprintnumbers, showpacs, nofootinbib, 
twocolumn]{revtex4-1}

\pdfoutput=1
\usepackage{amssymb}
\usepackage{amsthm}
\usepackage{graphicx,amsmath}
\usepackage{color}
\newcommand{\beq}{\begin{eqnarray}}
\newcommand{\eeq}{\end{eqnarray}}

\newcommand{\bmp}{\noindent\begin{minipage}{16cm}}
\newcommand{\emp}{\end{minipage}\vskip 7mm} 

\usepackage{dcolumn}
\usepackage{bm}
\usepackage{bbm}
\usepackage{subfigure}
\usepackage{slashed} 
\usepackage{mathrsfs}

\usepackage{epsfig}

\usepackage{hyperref}

\newcommand{\bea}{\begin{eqnarray}}
\newcommand{\eea}{\end{eqnarray}}

\newcommand{\ba}{\begin{eqnarray}}
\newcommand{\ea}{\end{eqnarray}}

       \newcommand{\nuDM}{\text{S}\nu \text{DM}}   
       
\newcommand{\be}{\begin{eqnarray}}
\newcommand{\ee}{\end{eqnarray}}

\begin{document}

\title{Neutrino Portal Dark Matter: From Dwarf Galaxies to IceCube}
%

\author{John F. Cherry}
\email{jcherry@lanl.gov}
\author{Alexander Friedland}
\email{friedland@lanl.gov}
\affiliation{Theoretical Division, Los Alamos National Laboratory, Los Alamos, New Mexico 87545, USA}

\author{Ian M. Shoemaker}
\email{shoemaker@cp3-origins.net} 
\affiliation{CP$^{3}$-Origins \& Danish Institute for Advanced Study  DIAS, University of Southern Denmark, Campusvej 55, DK-5230 Odense M, Denmark}


\begin{abstract}
It has been suggested that the baseline scenario of collisionless cold dark matter over-predicts the numbers of satellite galaxies, as well as the dark matter (DM) densities in galactic centers. 
This apparent lack of structure at small scales can be accounted for if one postulates neutrino-DM and DM-DM interactions mediated by light $\mathcal{O}({\rm MeV})$ force carriers. 
In this letter, we consider a simple, consistent model of neutrinophilic DM with these features where DM and a ``secluded'' SM-singlet neutrino species are charged under a new $U(1)$ gauge symmetry. An important ingredient of this model is that the secluded sector couples to the Standard Model fields only through neutrino mixing. We observe that the secluded and active neutrinos recouple, leading to a large relic secluded neutrino population.  This relic population can prevent small-scale halos from collapsing, while at same time significantly modifying the optical depth of ultra-high-energy neutrinos recently observed at Icecube. We find that the bulk of the parameter space accommodating an (a)symmetric thermal relic has potentially observable consequences for the IceCube high energy signal, with some of the parameter ranges already ruled out by the existing data. Future data may confirm this mechanism if either spectral absorption features or  correlations with nearby sources are observed.
 \end{abstract}

\preprint{CP3-Origins-2014-034 DNRF90, DIAS-2014-34, LA-UR-14-28339}

\maketitle

\section{Introduction}



There is a long-standing debate whether the collisionless cold dark matter (DM) predicts too much structure at small scales.  The conflict with observations is suggested in at least three ways: (1) {\it The cusp-versus-core problem} is the disagreement between the cuspy density profiles predicted from numerical simulations of collisionless cold DM and the cored profiles favored by well-studied dwarf galaxies~\cite{Moore:1994yx,Flores:1994gz,Navarro:1996gj}; (2) {\it the too-big-to-fail problem} is the observation that the most massive subhaloes in CDM simulations are too massive to host the satellites of the Milky Way, yet should be luminous given the observed dwarfs~\cite{BoylanKolchin:2011de,Walker:2012td}; and {(3)} the long-standing {\it missing satellites problem} {wherein} the number of satellites found in simulations of Milky Way sized halos disagrees with observations by roughly a factor $\mathcal{O}(10)$~\cite{Klypin:1999uc,Moore:1999nt,Kauffmann:1993gv}. While improved numerical simulations, incorporating baryons, and better satellite statistics are clearly needed and may yet alleviate this problem, the alternative possibility -- involving additional physics in the dark matter sector -- has justifiably received considerable attention. 

Particularly well-explored is the suggestion that DM could be self-interacting~\cite{Spergel:1999mh,Feng:2009mn,Loeb:2010gj,Tulin:2013teo,Fan:2013yva,Pearce:2013ola,Bellazzini:2013foa,Kahlhoefer:2013dca,Cyr-Racine:2013fsa,Bramante:2013nma,Kaplinghat:2013yxa,Cline:2013pca,Laha:2013gva,Curtin:2013qsa,Cline:2013zca,Boddy:2014yra,Hochberg:2014dra,Petraki:2014uza,Ko:2014bka,Wang:2014kja,Curtin:2014afa,Buckley:2014hja,Schutz:2014nka,Boddy:2014qxa}.
In this framework, the cusped centers of DM halos are smoothed into cores by providing DM with an efficient mechanism for transporting energy from the hot inner region to the cold outer region.  A velocity-dependent cross section can nicely accommodate the desire to have significant scattering at dwarf scales without running afoul of the stringent limits on self-interactions coming from cluster and galactic scales. 
Particle physics models with this feature employ $\mathcal{O}({\rm MeV})$ dark force mediators~\cite{Feng:2009mn,Loeb:2010gj,Aarssen:2012fx,Tulin:2013teo}. 

Additional benefits can be obtained if the hidden sector interaction could somehow also include neutrinos~\cite{Aarssen:2012fx}.
It has been argued that this scenario may solve \emph{all} of the small-scale structure issues mentioned above.
Indeed, the efficient scattering of DM with energetic bath particles would lead to late kinetic decoupling, delaying the formation of the smallest protohalos and resolving the problem of missing satellites~\cite{Aarssen:2012fx,Shoemaker:2013tda,Dasgupta:2013zpn}. For this process to damp structure on the scales relevant for the missing satellites problem, decoupling temperatures of order $\mathcal{O}(0.1~{\rm keV})$ are needed, singling out neutrinos as a potential scattering partner for DM. (though other realizations are possible~\cite{Chu:2014lja,Archidiacono:2014nda}). 

From the particle-physics point of view, the immediate issue to be addressed is how neutrinos could be consistently coupled to a hidden gauge group with the necessary strength without running into a host of constraints, e.g.,~\cite{Laha:2013xua}.
A straightforward path is use instead of the three Standard Model (SM) neutrinos a new, light fermion, which is a SM singlet~\cite{Bringmann:2013vra}. Provided this new \emph{secluded neutrino} could be produced in the early universe with sufficient relic abundance, the desired late kinetic decoupling of DM could be realized. 

\begin{figure*} 
\begin{center}
 \includegraphics[width=.2\textwidth]{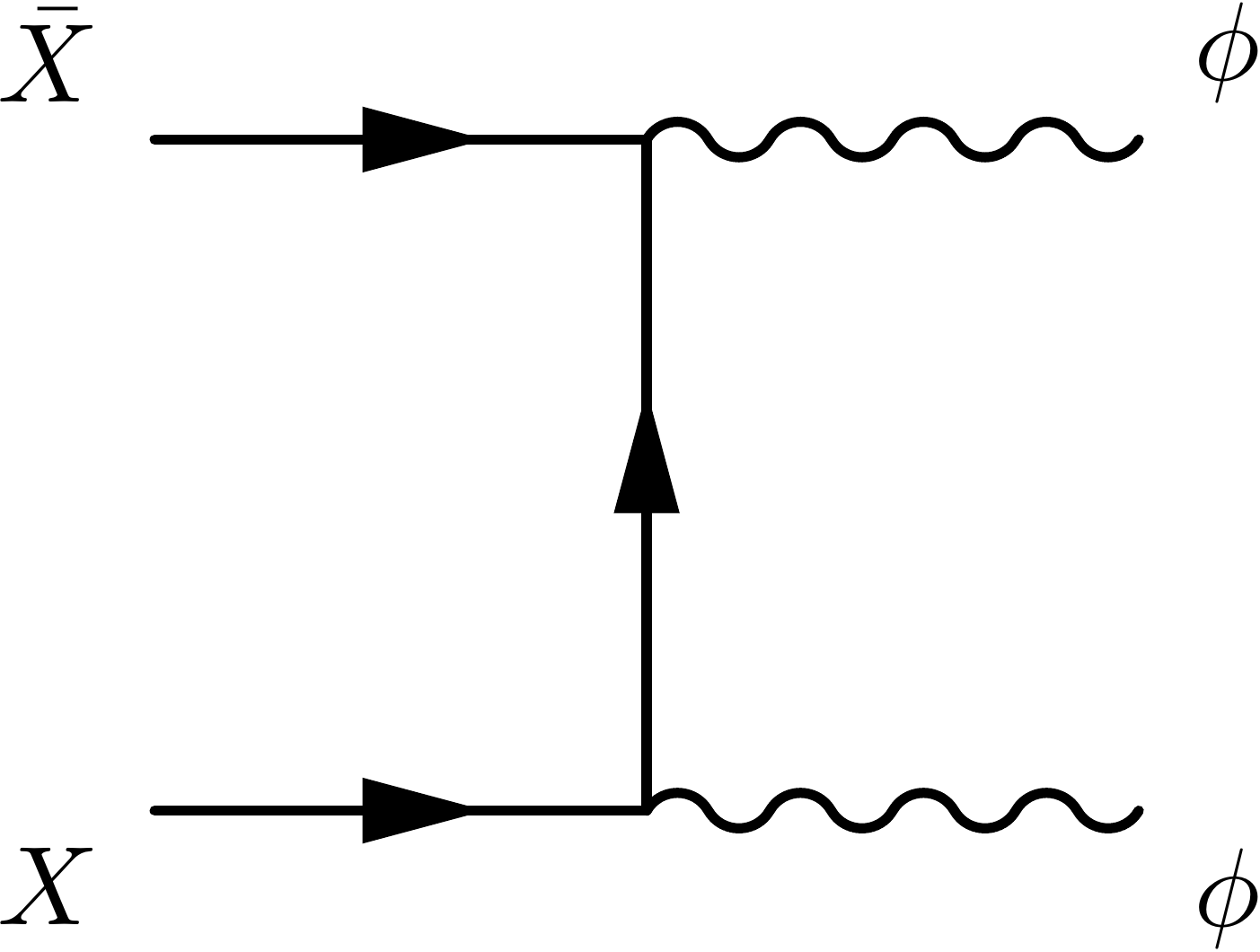}  ~~~~~~~~~~
\includegraphics[width=.2\textwidth]{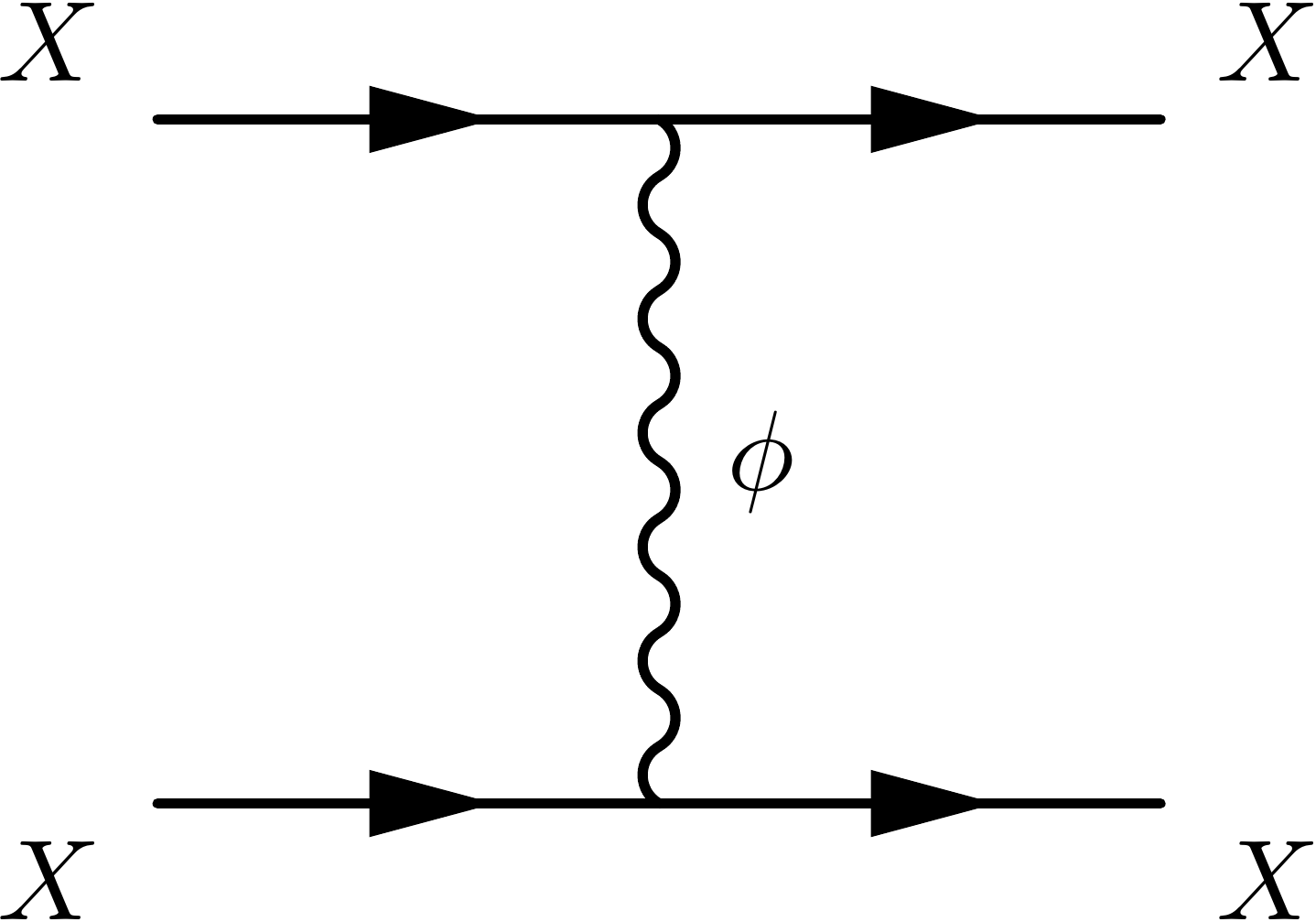}  ~~~~~~~~~~
\includegraphics[width=.2\textwidth]{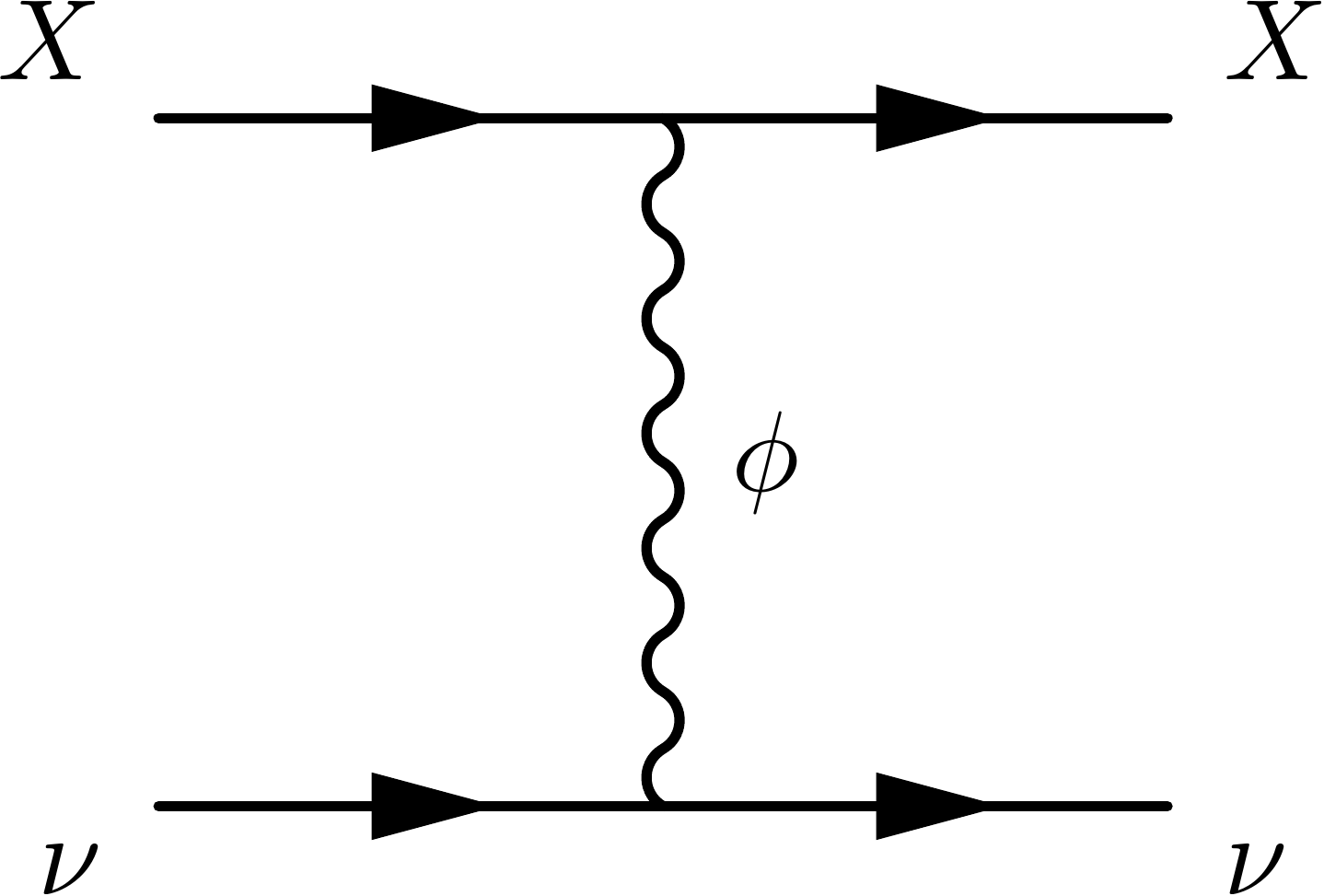}

\caption{ The relevant Feynman diagrams for (1) the relic abundance, (2) DM-DM self-scattering, and (3) $\nu$-DM scattering relevant for addressing the missing satellites problem.}
\label{fig:feyn}
\end{center}
\end{figure*}

The phenomenology of this secluded neutrinophilic DM ($\nuDM$) model contains several unique and salutary consequences, which are studied in this letter. First, we show the thermal recoupling of the active and {secluded} neutrino sectors indeed leads to a relic secluded neutrino population which is both hotter and more numerous than in previous studies~\cite{Dasgupta:2013zpn,Bringmann:2013vra}. Second, in contrast with~\cite{Aarssen:2012fx,Bringmann:2013vra} who considered symmetric DM we consider the larger parameter space in which DM carries a primordial particle-antiparticle asymmetry (though see~\cite{Dasgupta:2013zpn}). This has the effect of opening an interesting window in parameter space at low DM mass where self-scattering is in the perturbative regime. 
Lastly and most importantly, we point out that, thank to the active-secluded neutrino mixing, the neutrino self-interactions in this model modify the mean free path of Ultra-High-Energy (UHE) neutrinos as they propagate through the bath of relic neutrinos. We find that the bulk of the parameter space which simultaneously resolves all dark matter structure problems has direct observational consequences for the IceCube experiment.
\\
\indent
In Sec.~\ref{sec:model} we describe the general features of the {S}$\nu$DM model. In Sec~\ref{sec:relic} we solve the Boltzmann equations to determine the region of parameter space favored by an (a)symmetric thermal relic. In Sec.~\ref{sec:sidm} we determine the self-scattering parameters relevant for addressing the cusp-versus-core and too-big-to-fail problems. The {secluded} neutrino temperature and kinetic decoupling computation are addressed in Secs.~\ref{sec:steriletemp} and \ref{sec:KD} respectively. Implications of the neutrinos self-interactions for the high-energy IceCube data are discussed in Sec.~\ref{sec:icecube}. We summarize all of our results in Sec.~\ref{sec:discuss} and conclude.

\section{The $\nuDM$ Model }
\label{sec:model}

%

As already mentioned, the fact that simulations imply much cuspier density profiles than the cored profiles favored by observations~\cite{Moore:1994yx,Flores:1994gz,Navarro:1996gj} could be an indication that DM has non-negligible self-scattering~\cite{Spergel:1999mh}. 
Detailed analysis shows that a velocity-dependent interaction is favored, as can be achieved with a light force carrier. The argument proceeds as follows. The strongest constraints on DM self-interactions come from Milky Way ellipticity and Cluster collisions, roughly requiring $\sigma_{XX} /m_{X}\lesssim (0.1 -1)~{\rm cm}^{2}{\rm s}^{-1}$~\cite{Rocha:2012jg,Vogelsberger:2012ku,Peter:2012jh}. Note that these constraints are obtained from DM populations where the velocity dispersion is $\mathcal{O}(100~{\rm km/s})$ for Milky Way constraints and $\mathcal{O}(1000~{\rm km/s})$ for cluster constraints. For the $\mathcal{O}(1~{\rm cm}^{2}~{\rm g}^{-1})$ cross sections at dwarf scales ($\mathcal{O}(10~{\rm km/s})$), as identified by Spergel and Steinhardt~\cite{Spergel:1999mh}, to be allowed, the self-scattering should exhibits strong velocity dependence. Long-range interactions mediated by an $\mathcal{O}({\rm MeV})$ force carrier have precisely this feature and may thus solve the cusp-versus-core problem while remaining consistent with the constraints from galactic and cluster scales~\cite{Feng:2009mn,Loeb:2010gj,Aarssen:2012fx,Tulin:2013teo}.

{In this paper} we shall assume that the DM is a Dirac fermion, $X$, charged under a new $U(1)_{X}$ gauge interaction. There are two crucial ingredients for $\nuDM$, $\mathscr{L}_{\text{S}\nu\text{DM}} = \Delta \mathscr{L}_{\phi} + \Delta \mathscr{L}_{M}$, where the first term specifies the nature of the DM and neutrino coupling to the new gauge boson,
\be 
\Delta \mathscr{L}_{\phi} = g_{\nu} \bar\nu_{s} \gamma_{\mu}\nu_{s} \phi^{\mu} + g_{X} \bar{X} \gamma_{\mu}X \phi^{\mu}, 
\ee
and the second term,
\be 
\label{eq:Weinberg}
\Delta \mathscr{L}_{M} = 
y_{\alpha} \frac{(L_{\alpha} H)(h_{X} \nu_{s})}{\Lambda},
\ee
allows the new $\nu_{s}$ to mass-mix with the active SM neutrinos in a gauge-invariant way via a $U(1)_{X}$ charged Higgs $h_{X}$ which acquires a {vacuum expectation value (VEV)}. This Higgs is also responsible for giving mass to the vector, $m_{\phi} = g_{h}~ \langle h_{X} \rangle$, where $g_{h}$ is the gauge charge of the Higgs and $\langle h_{X} \rangle$ is its VEV.  Note that the active neutrinos are contained in their {electroweak (EW)} doublets, $L_{\alpha} = \left(
\begin{array}{c}
\nu_{\alpha}\\
\ell_{\alpha}\\
\end{array}
\right)$, 
where $\alpha = e, \mu, \tau$.

We note that the presence of this mixing is completely logical, since the operator in Eq.~(\ref{eq:Weinberg}) is suppressed by only a single power of the new physics scale $\Lambda$ and hence even new physics at very high scales could generate it. The situation is completely analogous to the standard Weinberg operator for the neutrino Majorana mass. Indeed, a simple ultraviolet completion of our model involves a see-saw type construction. One introduces right-handed singlet neutrinos with very large Majorana masses, couples them to both the SM and secluded neutrinos with Dirac mass terms and then integrates the heavy right-handed states out, yielding Eq.~(\ref{eq:Weinberg}) at low energies. 

The baryonic neutrino model of Pospelov~\cite{Pospelov:2011ha,Pospelov:2012gm,Pospelov:2013rha} employs similar features in order to endow neutrinos with new BSM interactions. We, however, do not assume any novel neutrino-baryon or neutrino-charged-lepton coupling. In fact, in $\nuDM$ when the universe is at temperatures below the high energy scale $\Lambda$, interactions between the dark and SM sectors can be mediated \emph{exclusively} through neutrino mixing. In this case, neither the ``dark photon'' searches nor DM direct detection experiments are expected to turn up a positive signal. The astrophysical and cosmological signatures discussed below, including the possible imprints of the dark sector in DM profiles and the Icecube data, become the \emph{primary} methods of discovering the dark sector

{Additional features of the model and the constraints on it are discussed in a concurrent publication~\cite{us2}. }


%

\section{Dark Matter Thermal Relic Abundance}
\label{sec:relic}
{
In models like ours where DM is not self-conjugate, there exists the possibility that some high-scale physics has violated the Sakharov conditions~\cite{Sakharov:1967dj} and generated a primordial asymmetry.  {This greatly expands the potential dynamic range of our model, as in our case the final abundance depends on both the annihilation cross section and the primordial asymmetry instead of the annihilation cross section alone.}  To be as general as possible, and since we are interested in relatively low-energy physics {in this paper}, we do not specify the nature of the high-energy physics producing this asymmetry (though see e.g.~\cite{Petraki:2013wwa,Zurek:2013wia} for examples), but instead simply impose the relevant annihilation requirements for an asymmetric thermal relic~\cite{Graesser:2011wi}.}

To find the requisite annihilation cross section at a given DM mass we solve the coupled Boltzmann equations, 
\be \frac{dn_{i}}{dt} + 3 H n_{i} = - \langle \sigma_{ann} v_{rel} \rangle \left[ n_{i}n_{j} - n_{eq}^2\right],
\label{eq:boltz}
\ee
where the indices run over $i,j = X, \overline{X}$, $H$ is the Hubble expansion rate, $n_{eq}(T)$ is the equilibrium number density, and $\langle \sigma_{ann} v_{rel} \rangle$ is the thermal average of the total annihilation cross section. 


We find that when the asymmetry is zero for Dirac DM, the correct abundance is obtained for $\langle \sigma_{ann} v_{rel} \rangle \simeq 4.5 \times 10^{-26}~{\rm cm}^{3}~{\rm s}^{-1}$ DM masses $\gtrsim 10$ GeV~\cite{Steigman:2012nb}. More generally however, when the asymmetry is nonzero, for the combination of the asymmetric and symmetric components not to overclose the Universe the annihilation cross section must be $\gtrsim 4.5 \times 10^{-26}~{\rm cm}^{3}~{\rm s}^{-1}$~\cite{Graesser:2011wi,Bell:2014xta}.  In what follows we will allow for nonzero asymmetry between $X$ and $\bar{X}$, and employ this constraint. 

Two processes contribute to the total annihilation cross section: $\overline{X}X \rightarrow \overline{\nu}\nu$ and $\overline{X}X \rightarrow \phi \phi$. When the DM mass is large compared to the mediator $\phi$ ($m_{X} > m_{\phi}$) the annihilation is governed by the diagram in Fig.~\ref{fig:feyn}(a), with cross section
\be 
\langle \sigma_{X\overline{X} \rightarrow \phi \phi} v_{rel} \rangle \simeq \frac{g_{X}^{4}}{16 \pi m_{X}^{2}}\sqrt{1-\left(\frac{m_{\phi}} {m_{X}} \right)^{2}},
\label{eq:2phi}
\ee
while the cross section for the $s$-channel annihilation to a pair of neutrinos is
\be
\langle \sigma_{X\overline{X} \rightarrow \nu \nu} v_{rel} \rangle \simeq \frac{N_{\nu} g_X^2g_{\nu}^{2}}{32\pi m_{X}^2}\frac{\left(2+\frac{m_\nu^2}{m_X^2}\right)}{\left(1-\frac{m_{\phi}^2}{4m_X^2}\right)^2+\frac{\Gamma_{\phi}^2}{m_{\phi}^2}}.
\ee
Thus whenever in the light mediator regime, $m_{X} > m_{\phi}$, the $\overline{X}X \rightarrow \phi \phi$ channel dominates so long as $g_{X} > g_{\nu}\sqrt{N_{\nu}}$.

With the assumption that the $\overline{X}X \rightarrow \phi \phi$  mode dominates, this effectively fixes the value of $g_{X}$ in the symmetric DM limit or serves as a lower limit on $g_{X}$ when there exists a nonzero particle asymmetry. 
Therefore using Eq.~(\ref{eq:2phi}) and imposing that the annihilation cross section be $\gtrsim 4.5 \times 10^{-26}~{\rm cm}^{3}~{\rm s}^{-1}$ , we find that an (a)symmetric thermal relic roughly requires that the DM coupling is, $g_{X} \simeq 0.02~ \sqrt{m_{X}/{\rm GeV}}$. 

{The above discussion is modified at large DM masses by the presence of Sommerfeld-enhanced scattering~\cite{Hisano:2004ds,ArkaniHamed:2008qn,Feng:2010zp,Tulin:2013teo}, though is largely irrelevant for the sub-TeV DM masses with which we are concerned here~\cite{Tulin:2013teo}.}


\section{DM Self-Scattering: turning cusps into cores}
\label{sec:sidm}
{
As discussed in Sec.~\ref{sec:model} dwarf galaxies may indicate the need for DM-interactions, while cluster and galactic observations only yield limits on DM self-interactions. Therefore, along the lines of~\cite{Loeb:2010gj,Tulin:2013teo,Aarssen:2012fx} we are interested in exploring the parameter space of a thermal relic for which the DM-DM scattering cross section is large at dwarf speeds but small enough to accommodate galactic and cluster limits on self-interactions~\footnote{Though we stress that velocity-independent interactions around $\sim$0.5$~{\rm cm}^{2}{\rm g}^{-1}$ may be sufficient to modify the profiles of dwarfs while narrowly evading galactic and cluster scale limits~\cite{Rocha:2012jg,Peter:2012jh}}. Given this set of constraints, we adopt a region of interest for self-interactions defined as~\cite{Tulin:2013teo}: 1-10$~{\rm cm}^{2}{\rm g}^{-1}$ at characteristic dwarf speeds, $10~{\rm km}/{\rm s}$, while $< 1~{\rm cm}^{2}{\rm g}^{-1}$ at galactic and cluster speeds, $100-1000~{\rm km}/{\rm s}$ respectively. }

In an asymmetric DM context as we are considering, the parameter space for self-interactions is quite large given that relic abundance considerations only impose a lower bound on the coupling, $g_{X}$, rather than fixing it as in the symmetric relic case. In contrast with~\cite{Aarssen:2012fx,Bringmann:2013vra}, who considered symmetric DM, we shall see that thermal asymmetric DM allows for a potential resolution of the small-structure problems in the perturbative regime as well. 

{In the small-coupling regime ($\alpha_{X}m_{X} / m_{\phi} \ll 1$) where the scattering can be computed perturbatively, one can use the Born approximation to compute the $t$-channel contribution to the transfer cross section~\cite{Feng:2009hw,Tulin:2013teo}. We agree with this calculation and find,}
\be 
\sigma_{T} = \frac{g_{X}^4}{2\pi m_{X}^{2} v^{4}} \left[\ln \left(1+ \frac{m_{X}^{2}v^{2}}{m_{\phi}^{2}}\right) - \frac{m_{X}^{2} v^{2} }{m_{\phi}^{2}+m_{X}^{2}v^{2}}\right].
\ee

{In the non-perturbative regime ($\alpha_{X}m_{X} / m_{\phi} \gg 1$), fitting expressions have been obtained in the limit that the DM de Broglie wavelength is small compared to the characteristic scale of the potential, $m_{X}v/m_{\phi} >1$. In this sense the scattering proceeds ``classically.'' For repulsive scattering these have been found to be \cite{Khrapak,Tulin:2013teo},}
\begin{equation}
\sigma_T \approx \left\{ \begin{array}{cc} 
\frac{2\pi}{m^2_\phi}& \beta^2 \ln(1+\beta^{-2}),
 \beta<1, \\
\frac{\pi}{m^2_\phi}&
\left( \ln 2\beta - \ln \ln 2 \beta \right)^2,
\beta > 1,
\end{array} \right.
\end{equation}
{where $\beta \equiv 2 \alpha_X m_\phi / (m_\chi v_\mathrm{rel}^2)$ is the ratio of the potential energy at the characteristic length scale $r = m_{\phi}^{-1}$ to the kinetic energy of the DM.}

Outside the realm of applicability for either of these analytic results, we solve the Schr$\ddot{{\rm o}}$dinger equation using the numerical recipe outlined in~\cite{Tulin:2013teo}.

\section{{Secluded} Neutrino Abundance and Temperature}
\label{sec:steriletemp}

After the {{d}ark sector decouples from the Standard Model, the temperature ratio between radiation in the two sectors is easily estimated from the conservation of entropy under the assumption that the two sectors shared a common temperature in the past, $T_{d}$. This is found to be
\be \left. \frac{T_{s}}{T_{\gamma} }\right |_{T_{KD}} = \left[ \frac{ g_{*,s}(T_{d})~~g_{*,SM}(T_{KD})}{g_{*,SM}(T_{d})~~g_{*,s}(T_{KD})}\right]^{1/3}\, ,
\ee
where $T_{KD}$ is the temperature of kinetic decoupling (discussed in the next section) and $T_{s}$ is the temperature of the secluded neutrinos.  Thus taking $T_{d} = 1$ TeV and assuming both the scalar and vector were in equilibrium at early times but not at kinetic decoupling, we find $T_{s}/T_{\gamma} \simeq 0.47$. 

{An important observation from} models of neutrinos with dark sector interactions is that immediately after decoupling the one-loop finite temperature self-energy contribution to the neutrino effective mass strongly suppresses the mixing angle between the active and {secluded} neutrinos~\cite{Hannestad:2013ana,Dasgupta:2013zpn}.  {The effect is the direct analog of the standard model finite temperature potential for neutrinos~\cite{Notzold:1988fv}.  This mixing angle suppression isolates} the dark sector from the standard model sector by reducing the rate of $\nu_{\rm active} - \nu_{\rm {s}}$ scattering, to the extent that it is much less than the expansion rate of the universe.  This prevents {the secluded sector} from thermalizing with the Standard Model sector through neutrino scattering at temperatures above $15\,\rm MeV$.  {As a result, the phenomenology of Big Bang Nucleosynthesis (BBN) is unaffected by the addition of secluded neutrino interactions.}

{A critical feature of the model we present in this paper which was missed by previous authors~\cite{Hannestad:2013ana,Dasgupta:2013zpn} is the recoupling of the secluded neutrinos to the active neutrino population at low (post BBN) temperatures.  Once the temperature of the relic {secluded} neutrinos has dropped to a sufficiently low level, the thermal contribution to the {secluded} neutrino self energy will become small enough that it no longer suppresses the mixing angle between active and {secluded} neutrinos.  Once the mixing angle becomes sufficiently large, the rate for 2 to 2 neutrino scattering through the dark sector interaction becomes fast enough to equilibrate the populations of secluded and active neutrinos.}  This occurs roughly when the neutrino oscillation rate is comparable to the effective potential,
\be 
\frac{\delta m^2}{2E_s}\cos 2\theta_s \simeq \frac{7\pi^2 g_\nu^2 E_s T_{s}^4}{45 m_\phi^4}\, , 
\ee
where we have employed the form of the effective potential valid at low-temperatures, $T_{s}, E_{s} \ll m_{\phi}$~\cite{Notzold:1988fv,Dasgupta:2013zpn}. 

{A number of neutrino beam experiment anomalies provide a guide for our choice of neutrino mixing parameters~\cite{Aguilar:2001pj,Aguilar:2013ik}, which we take to be $\delta m^2 = 1 {\rm eV}^2$ and $\theta_s = 0.1$.  This has the benefit of placing our neutrino mixing portal in a regime that future short baseline neutrino experiments may be able to probe.  For the range of $g_s$ and $m_\phi$ which satisfy the constraints placed on SIDM, along with the {secluded} neutrino mixing parameters above}, we find that mixing angle suppression will cease in the temperature range $500\,{\rm keV} - 5\,{\rm keV}$.  Without mixing angle suppression, the rate of scattering between active and {secluded} neutrinos through the dark sector interactions becomes much faster than the Hubble expansion rate.

Interestingly, the newly reconnected neutrino populations do not truly thermalize with each other.  The temperature of decoupling is much lower than $2m_\phi$ so that neutrino number changing interactions are not possible.  Further, recoupling occurs below the temperature at which the active neutrinos are decoupled from the SM plasma.  Because no new entropy can enter or be generated within the dark-active coupled neutrino system, this leads to an equilibrium state which is dictated by detailed balance.  The scattering processes which convert {secluded} neutrinos to active neutrinos and vice versa are both of the same magnitude in equilibrium, $\mathcal{O}\left( \sin^2\theta_s\right)$, so that the final state of the system is a sum of the Fermi-Dirac distributions of the {secluded} and active neutrinos weighted by the relative number of degrees of freedom in the {secluded} and active sectors.  Though this process equilibrates the effective temperatures of all neutrinos, the spectrum of {secluded} (and active) neutrinos is distorted in such a way that a fractional value of a fully thermally populated neutrino species remains. We find that to a good approximation, the scattering processes yield a detailed balance with {$84\%$ the number density of a fully thermalized neutrino distribution} and $T_{s} \simeq T_{\nu} = (4/11)^{1/3} T_\gamma$, for our initial condition of a single {secluded} neutrino species which decouples at $T_{d} = 1\,\rm TeV$. 

{This change in the effective temperature of the secluded neutrinos is a key factor in computing the correct kinetic decoupling temperature, $T_{KD}$, which we perform in the next section.  As we will shortly see in Eqn.~\ref{T_KD}, the increased temperature of the secluded neutrinos is more impactful on $T_{KD}$ than the fractional thermal population engendered by detailed balance.  As a result, our model predicts the effect of secluded neutrino scattering on the small scale structure of DM halos to be much more pronounced than previously calculated~\cite{Aarssen:2012fx,Dasgupta:2013zpn}.}

\section{Late kinetic decoupling: where the missing satellites went}
\label{sec:KD}

Even after the number-changing annihilation reactions cease being in equilibrium, kinetic equilibrium between DM and the {secluded neutrinos} can persist via elastic scattering, $X \nu_{s} \leftrightarrow X \nu_{s}$. This late kinetic decoupling delays the formation of the smallest protohalos and may offer a solution to the missing satellites problem~\cite{Aarssen:2012fx,Bringmann:2013vra,Dasgupta:2013zpn,Ko:2014bka,Chu:2014lja}. 

\begin{figure*} 
\begin{center}
    \includegraphics[width=.45\textwidth]{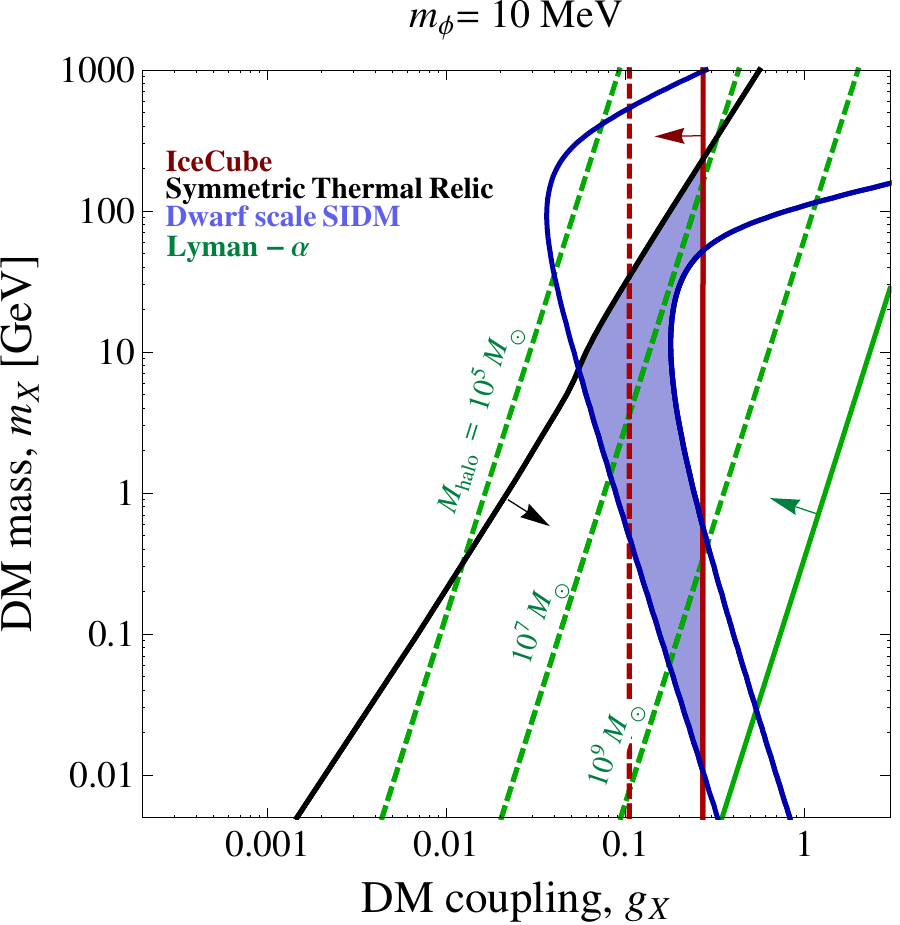} 
      \includegraphics[width=.45\textwidth]{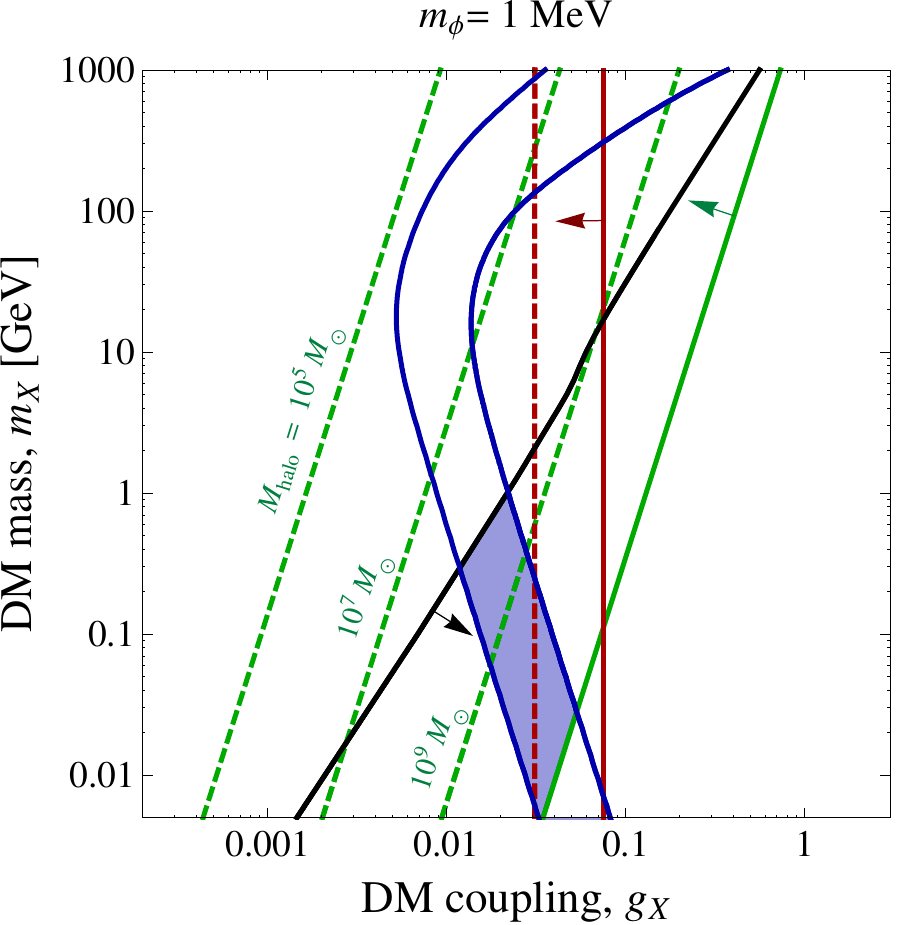} 
\caption{ {In the {\it left} and {\it right} panels we fix the mediator mass to 10 MeV and 1 MeV respectively. To address the cusp-versus-cores and too-big-to-fail problems a parameter point should lie within the shaded blue region. Values to the left of the black curve lead are excluded by producing an over-abundance of DM, $\Omega_{DM} h^{2} > 0.12$. Regions to the right of the green curve are excluded by Lyman-$\alpha$ requirements that $M_{halo} < 5 \times 10^{10}~M_{\odot}$, {while regions to the right of the red solid line are excluded by having a MFP $<$ 50 Mpc. In the region to the right of the dashed red IceCube can perform source correlations at 3$\sigma$ (see text for details).} For reference the dashed green curves are contours of constant $M_{halo} = 10^{5}~M_{\odot},10^{7}~M_{\odot}, 10^{9}~M_{\odot}$ from left to right. Arrows indicate the direction in which the parameter space is allowed.}  }
\label{fig:gold}
\end{center}
\end{figure*}

The momentum relaxation rate from this process can be roughly estimated from, $\Gamma_{p} \sim \sigma_{X \nu}~ n_{d} ~\left(T_{KD}/m_{X}\right)$, where the factor in parentheses accounts for the fact that many neutrino scatterings are required to appreciably change the DM momentum. Then using $n_{s} = \frac{3 }{2} \frac{\zeta(3)}{\pi^{2}} T_{s}^{3}$, one can estimate the temperature of kinetic decoupling by equating the momentum relaxation rate to the Hubble rate, $\Gamma_{p} = H$. 


Though the above sketch is qualitatively correct, we employ a method along the lines of~\cite{Gondolo:2012vh} which incorporates the effects of Fermi-Dirac statistics and Pauli blocking. With this method we find the temperature of DM-{secluded} neutrino kinetic decoupling to be (see Appendix for details):
\bea 
T_{KD} =  \left(\frac{104.58}{31 \pi^{3} N_{\nu}}\right)^{1/4} \frac{m_{\phi} g_{*}^{1/8}}{\sqrt{g_{X}g_{\nu}}} \left(\frac{m_{X}}{M_{Pl}}\right)^{1/4}\left(\frac{T_{\gamma}}{T_{s}}\right)^{3/2}_{KD}.
\label{T_KD}  
\eea
where $g_{*}$ is the effective energy degrees of freedom parameter. 


{The final ingredient to make contact with the missing satellites problem is to relate the temperature of kinetic decoupling to the mass of the smallest DM protohalos. As observed in~\cite{Hofmann:2001bi,Loeb:2005pm,Bertschinger:2006nq}, as long as DM-SM interactions are in kinetic equilibrium acoustic oscillations damp structure on sub-horizon scales. After kinetic decoupling, DM can stream freely out of over-densities and wipe out structure up to sub-free streaming scales. The mass of the smallest protohalos is determined by the largest of these two effects. For the low values of kinetic decoupling we consider here acoustic damping is dominant~\cite{Bringmann:2009vf,Gondolo:2012vh,Cornell:2013rza}.}  Thus the concomitant suppression in the halo mass function is simply estimated from the amount of DM in the horizon at temperature $T_{KD}$~\cite{Bringmann:2009vf}
{
\bea 
M_{halo} &=& \frac{4\pi}{3} \rho_{DM,0}\frac{g_{*,s}(T_{KD})}{g_{*,s}(T_{0})}\left(\frac{T_{KD}}{T_{0}}\right)^{3}H^{-3}(T_{KD}) \nonumber \\
&=&1.7 \times 10^{8}~M_{\odot} \left(\frac{{\rm keV}}{T_{KD}}\right)^{3} 
\label{M_halo}
\eea
where $g_{*,s}$ is the entropy degrees of freedom, $T_{0}$ is the temperature of the present epoch, $\rho_{DM,0}$ is the present DM density and, $M_{\odot} \simeq 1.1 \times 10^{57}$ GeV is a solar mass.} For halo cutoffs addressing the missing satellites problem, $10^{9-10}M_{\odot}$, this requires temperatures $T_{KD} \simeq 0.1-0.5$ keV. 


\section{Implications for IceCube}
\label{sec:icecube}

\subsection{Basic argument}

We now come to the crucial stage of our presentation. The basic observation for what follows is that, if the dark force mediates DM-DM and DM-$\nu_{s}$ interactions, it should also mediate \emph{self-interactions of $\nu_{s}$}. If the secluded and active neutrinos appreciably mix, as in Eq.~(\ref{eq:Weinberg}), all neutrino mass eigenstates are endowed with this novel self-interaction. The result is anomalous scattering in collisions of neutrino particles. While a requisite laboratory neutrino-neutrino collider is presently lacking, the Universe provides an excellent setup for testing of this scenario, with baselines on gigaparsec scales. The ``beam'' is supplied by the ultra-high-energy (UHE) neutrino flux recently observed at the Icecube experiment; the background, by the relic populations of secluded -- and to a lesser extent active -- neutrinos.

Indeed, consider a neutrino produced at an astrophysical source. Initially in an active state $\nu_{a}$, it quickly oscillates into other flavors, and eventually separates into several wave packets corresponding to the different mass eigenstates that $\nu_{a}$ projects onto. Let us assume a generic scenario that the three mostly-active mass eigenstates, $\nu_{1,2,3}$,  each have a similar, small amount of the secluded admixture, $\sim\sin\theta_{s}$. The fourth state, $\nu_{4}$, is then mostly made up of secluded neutrino, $\sim\cos\theta_{s}$. The initial active state $\nu_{a}$ has a small probability, $\sin^{2}\theta_{s}$, of immediately projecting into $\nu_{4}$. This component essentially disappears from the flux, having only a probability of  $\sin^{2}\theta_{s}$ to interact as an active neutrino in the Icecube detector (see below). 

A more interesting fate awaits the mostly-active eigenstates, $\nu_{1,2,3}$. While propagating through the relic background of $\nu_{4}$, these UHE neutrinos are subject to \emph{flavor-dependent} interactions. Specifically, only the $\nu_{s}$ components of $\nu_{1,2,3}$ enter the interactions. The final state of the scattering then consists of a $\nu_{s}-\nu_{s}$ pair, with each most likely to project onto $\nu_{4}$ state. Thus, the combination of dark-force-mediated scattering and the mixing-induced oscillations effectively converts active neutrinos into secluded ones, depleting the UHE neutrino flux. 

Three important observations are in order. First, the above discussion assumes that the secluded neutrino is present at an appreciable level in all three states, $\nu_{1,2,3}$. If one or more of these states do not contain $\nu_{s}$, they are not subject to the scattering process, resulting in only partial suppression of the flux. Second, the active relic neutrino background also scatters the UHE flux, but with the probability suppressed by an additional factor of $\sin^{2}\theta_{s}$. Lastly, one may ask whether the active neutrino flux can be regenerated by subsequent scattering. After all, each subsequent event has a $\sim\sin^{2}\theta_{s}$ probability of producing $\nu_{1,2,3}$ and the $\nu_{4}$ neutrinos are subject to more frequent interactions with the relic background than in $\nu_{1,2,3}$, owing to the larger content of $\nu_{s}$. It is important to note, however, that such regenerated flux would have significantly lower energies, since each scattering event distributes the energy of the incident UHE neutrino between the two daughter states. Effectively, this flux disappears from the ultra-high-energy spectrum.

The efficiency of the scattering depends on the ratio of the mediator mass $m_{\phi}$ and the center-of-mass energy $s=2E_{\nu}m_{s}$, where $m_{s}$ is the mass of the mostly-secluded state $\nu_{4}$. Specifically, for a $t$-channel exchange, one has
\be
\sigma^t_{\nu\nu}(z) = \begin{cases} 
\sin^2\theta_s \frac{s g_s^4}{2\pi m_\phi^4},\ &s \ll m_\phi^2\, , \\
\sin^2\theta_s \frac{3g_s^4}{4\pi m_\phi^2},\ &s\gg m_\phi^2 \,.\end{cases}
\ee
Below the mediator mass, the interaction becomes effectively contact and, just like for the SM Fermi interaction, the strength decreases with decreasing energy. For strong absorption, we need to be in the regime $s \gtrsim m_{\phi}^{2}$. 

Remarkably, this is indeed realized for us. For  a $\sim$ {100 TeV} astrophysical neutrino and a $\sim$1 eV secluded neutrino mass, as motivated by the short-baseline anomalies~\cite{Aguilar:2001pj,Aguilar:2013ik}, the center-of-mass energy is {$\sim 10$ MeV}, which is \emph{exactly the scale of the mediator masses favored by the velocity-dependent DM self-scattering in galactic cores}. In this case, the $t$-channel cross section is {$\sim9300$ fm$^{2}$ (1 MeV$/m_{\phi})^{2} g_{s}^{4}\sin^{2}\theta_{s}$} and the corresponding
 mean free path (MFP)  assuming the relic number density of ${\cal O}(10^{2})$ cm$^{-3}$ is {$\sim 30$ pc $ (m_{\phi}/\mbox{1 MeV})^{2} g_{s}^{-4}\sin^{-2}\theta_{s}$}.
Thus, the UHE neutrinos at Icecube provide an excellent probe of our scenario.

In their recently released three-year dataset the IceCube collaboration has reported 37 events above the atmospheric neutrino background with energies between 30 and 2000 TeV, with a significance of 5.7$\sigma$~\cite{Aartsen:2014uq}.  The origin of these high-energy neutrino events remains unknown, though they appear to be isotropically distributed, suggesting an extragalactic origin.  If this is the case, the MFP of high-energy neutrinos as they scatter on the C$\nu$B cannot be too short, as most of the flux originating at cosmological distances would not reach us. This can be immediately seen from Fig.~\ref{fig:absorption}, which shows the  fraction of events at IceCube expected to originate within a given redshift, $z$, assuming SM neutrino interactions and the source distribution that tracks the star formation history of the universe.
Even if one boosts the flux emitted by the nearby sources by a large factor, the observed flux would look highly anisotropic. This consideration leads to an {\it upper bound} on the coupling in $\nuDM$. 

Indeed, taking $\sin^{2}\theta_{s}=0.01$ and demanding that the MFP be at least 50 Mpc for isotropy considerations, {we find $g_{s}\lesssim 0.09 (m_{\phi}/\mbox{1 MeV})^{1/2}$, or $\alpha_{s}=g_{s}^{2}/4\pi\lesssim 6\times 10^{-4} (m_{\phi}/\mbox{1 MeV})$,} a significant constraint. This constraints is illustrated in Fig.~\ref{fig:gold}: it is responsible for the right cutoff on the allowed  regions. Similar considerations were made for toy models~\cite{Ioka:2014kca,Ng:2014pca}.

For couplings below this cutoff, we have an interesting possibility of \emph{probing} the $\nuDM$ scenario with future Icecube data. There are at least two smoking-gun signatures to look for in this case. First is the effect of absorption in certain energy bands due to the $s$-channel resonance. The $s$-channel cross section is suppressed only by 2 powers of the coupling constant, being connected to the production of the physical particle in the process of $\nu_{s}\nu_{s}\rightarrow\phi$. Thanks to the red-shifting of the energy, the $s$-channel absorption could result in a gap in the energy spectrum. Such a gap, if confirmed with enough statistics, would be difficult to ascribe to physics at the source.

\begin{figure} 
 \includegraphics[width=.45\textwidth]{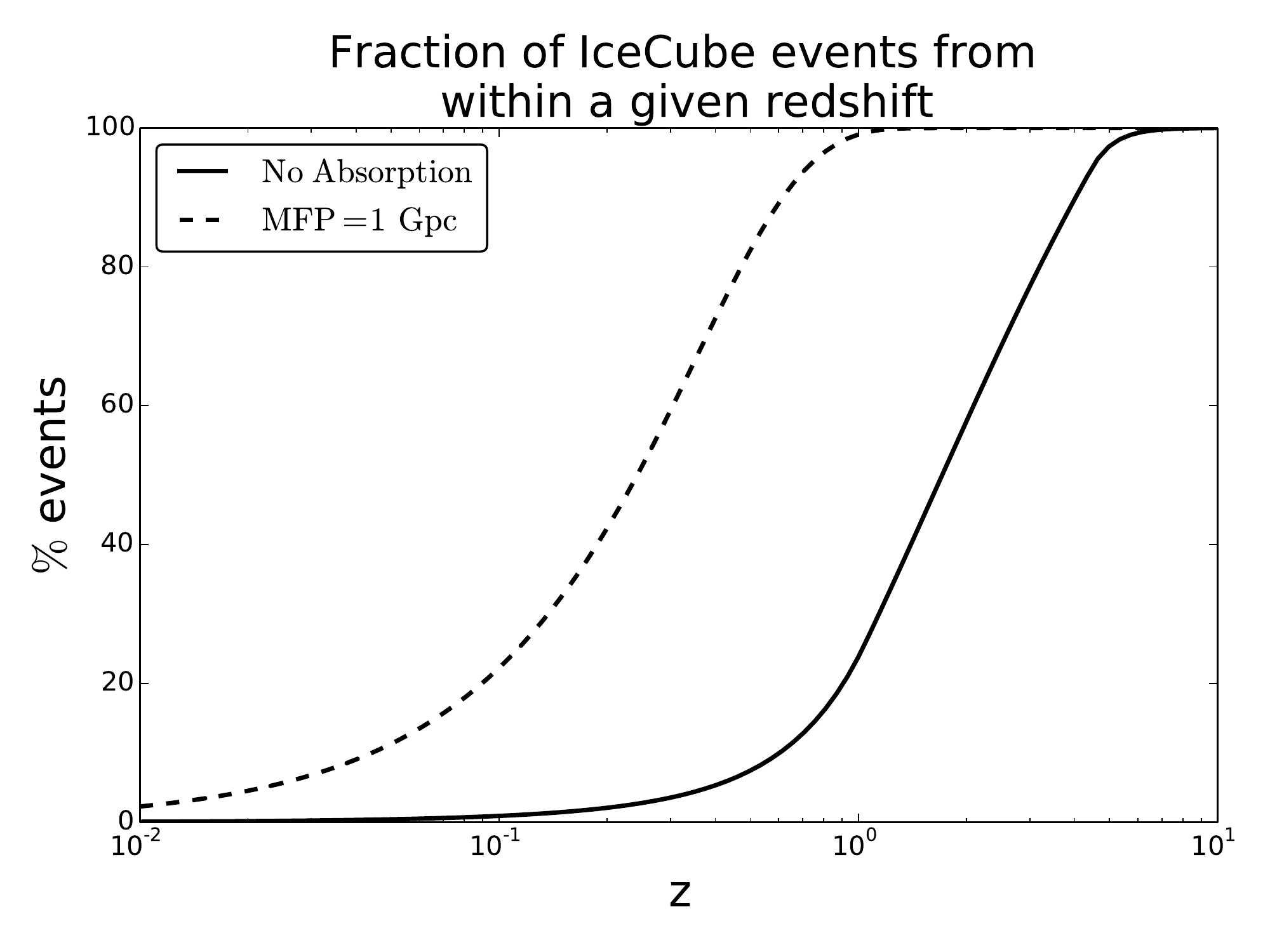} 
\caption{ Here we illustrate the effect of neutrino scattering on the fraction of events at IceCube originating within a given redshift, $z$. Notice that when absorption is present, a larger fraction of events originate from nearby, and may be more easily correlated with known sources.}
\label{fig:absorption}
\end{figure}

Second is the possibility that the observed UHE neutrinos could be correlated with known nearby sources. Indeed, correlations with distant sources is not expected, since most of sources at cosmological distances (redshift of $z\sim1-5$) are not in any catalog. If nearby source correlations were to be observed, one would conclude that a large fraction of the flux is \emph{missing}. The argument is that, on generic grounds, one  expects sources to follow a distribution similar to the star-formation history of the universe. Then, as seen in Fig.~\ref{fig:absorption}, the population at $z\sim1-5$ should contribute most of the flux. For example, if the observed neutrinos were to be correlated to sources lying within $z \lesssim 0.2$, distant sources would be expected to contribute some 50 times as much flux. Its absence would then imply \emph{a neutrino redshift horizon}, pointing towards the $\nuDM$ scenario. We next consider this scenario in some detail.


\subsection{Detailed example}

\begin{figure*} 
\begin{center}
 \includegraphics[width=.45\textwidth]{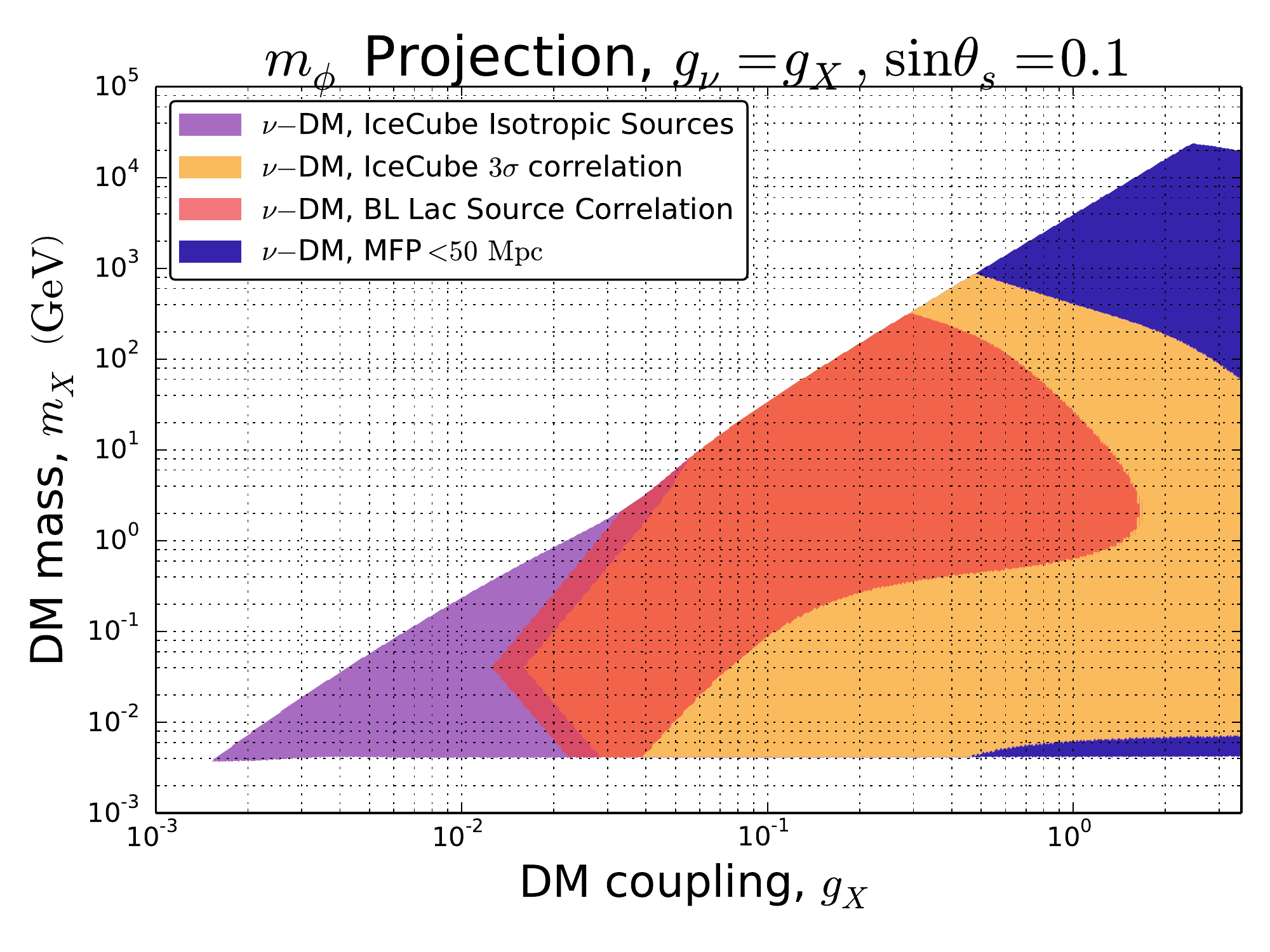} 
  \includegraphics[width=.45\textwidth]{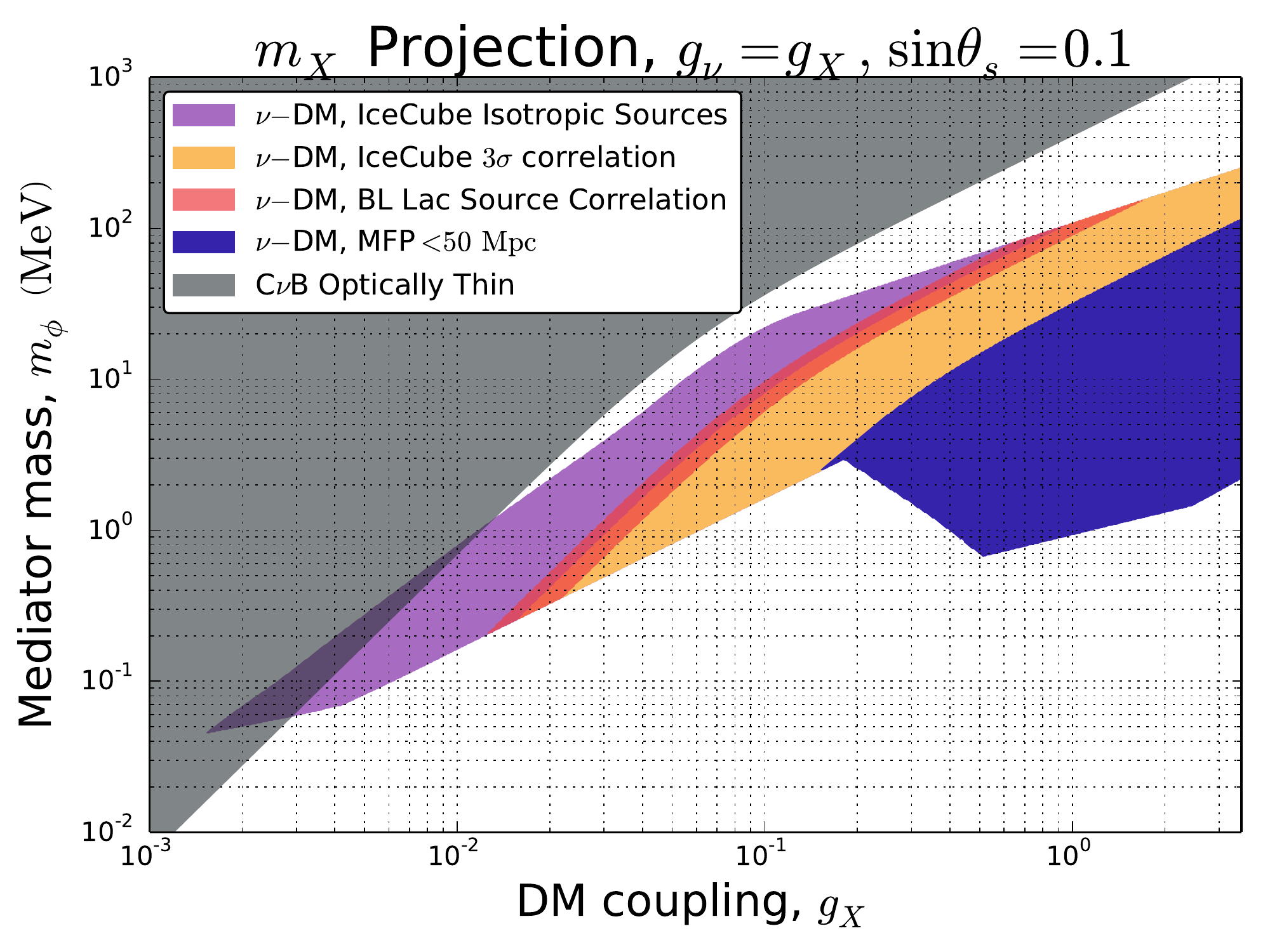}
\caption{  {\it Left}: 2D projection along the $m_\phi$ axis showing the regions of the $m_X$-$g_X$ parameter space which may potentially solve the dark matter structure problems while producing identifiable absorption features for the IceCube experiment. {\it Right}: 2D projection along the $m_X$ axis showing the regions of the $m_\phi$-$g_X$ parameter space which may potentially solve the dark matter structure problems while producing identifiable absorption features for the IceCube experiment.  {Blue indicates the regime where the C$\nu$B is opaque to high energy neutrinos on distances less than $50\,\rm Mpc$, orange color indicates regimes where the C$\nu$B is opaque to high energy neutrinos on distances short enough that absorption might be detected via IceCube source correlations at the level of $3\sigma$ statistical significance (using the 3 year data set~\cite{Aartsen:2014uq}), {{purple} indicates the regime where there absorption of high energy neutrinos may alter the IceCube observed spectra without creating a significant source correlation,} red indicates the regime where the absorption of high energy neutrinos reconciles the over abundance of IceCube events correlated with BL Lacs at $z < 0.212$, and dark grey regions show the regime where the C$\nu$B is optically thin out to $z=10$.}}
\label{fig:projections}
\end{center}
\end{figure*}

Let us now proceed to a more detailed examination of the modification of the neutrino optical depth in $\nuDM$. Assume a relic background of {secluded} neutrinos with a number density of $n_{\nu_s}\vert_{z=0} = 94\,{\rm cm}^{-3}$ for our example of $T_{d} = 1\,\rm TeV$. 
We will use the criterion that the optical depth 
for a high energy neutrino to scatter with the C$\nu$B be greater than unity.  We compute the optical depth as follows,
\be
\tau = \int_0^{r_p} n_{\nu_s}\left( z\right) \sigma_{\nu\nu}\left( z\right) dr_p = \int_0^{z_{i}} \frac{ c n_{\nu_s}\left( z\right) \sigma_{\nu\nu}\left( z\right) dz}{(1+z)H(z)}\, ,\nonumber \\
\ee
where $r_p$ is the proper distance along the neutrino world line, $z$ is the redshift, $H(z)$ is the Hubble expansion rate, and the number density of background {secluded neutrinos} $n_{\nu_s} = n_{\nu_s}\vert_{z=0} (1+z)^3$.  Contributions to $\sigma_{\nu\nu}\left( z\right)$ will come from resonant $s$-channel scattering and from $t$-channel scattering.  The resonant scattering, as explained above, could yield distinct absorption bands, which would be smoking-gun signatures of our mechanism. At the same time, the locations of these band  are {sensitively dependent on both} the mediator mass and the absolute neutrino mass scale. Hence,  {determining} whether or not resonant absorption features will appear in the IceCube data is not a reliable constraint.  Therefore we chose to use the $t$-channel scattering alone to constrain what portions of the parameter space may impact IceCube observations.  For concreteness, we compute the bounds for our different scattering regimes using $E_\nu(z=0) = 63\,\rm TeV$, which corresponds to the lowest IceCube event energy which has been claimed to correlate with a gamma ray source at known {redshift}~\cite{Padovani:2014uq}.} 

The first scattering regime we define is the {Optically Thin regime, where $\tau < 1$ out to a redshift of $z = 10$.  This regime will be unlikely to have an observable effect on the IceCube signal as the potential sources for TeV - PeV neutrinos (such as AGNs, GRBs, or star forming galaxies) have redshift distributions which typically peak at $z < 4$.  The second regime is the mean free path $< 50\,\rm Mpc$ regime, where the optical depth for neutrino absorption is $\tau = 1$ at a distance of $50\,\rm Mpc$ or less.  This leads to IceCube sources which can be directly correlated with local large scale structure around or within the Milky Way galaxy. 

{The third regime we define is the IceCube 3$\sigma$ correlation projected limit.  In this regime the absorption of high energy neutrinos is sufficiently strong that it is possible to use the 3 year data release from IceCube~\cite{Aartsen:2014uq} to detect a discrepancy between the  low {redshift} ($z \leq 0.2$) distribution of correlated IceCube sources and the redshift distribution of the same sources as identified with photons.  To compute this limit {we take the median number of events above background in the IceCube 3 year data set, 20, and assume that the {red}{redshift distribution of} IceCube neutrino sources tracks the star formation rate (which is likely for many potential sources~\cite{Aartsen:2014uq}).  This yields the expectation that $0.34$ IceCube neutrino events should originate within the range $0 < z < 0.2$.  We then compute the effects of absorption in our model on the redshift distribution of neutrinos emitted from sources distributed according to the star formation rate, taking the overall flux and {red}{adjusting the normalization} such that the IceCube experiment is expected to observe 20 events {\it post-absorption}.  The post-absorption expected number of events in the redshift range $0 < z < 0.2$ is then tallied and the significance of the discrepancy between the absorbed and non-absorbed expectations computed.  Requiring the discrepancy be at least $3\sigma$ confidence level or greater yields the constraints on $m_\phi$ and $g_\nu$ shown in Figure~\ref{fig:projections} as well as a the projection that the IceCube detectable redshift horizon (beyond which $\tau>1$) for $t$-channel absorption is $z \leq 0.70\,(0.64)$ for the contact (continuum) interaction limit.}}  
 
 The {fourth} regime we define is the Isotropic Source regime, where the optical depth for high energy neutrino absorption is greater than unity only for redshifts greater than $z=0.70\,(0.64)$ in the contact (continuum) interaction limit.  The absorption of high energy neutrinos in this regime may yet alter the spectral index of the high energy neutrino spectrum or create absorption lines, but it will not be detectable through correlating IceCube events with astrophysical sources.}

{The recent results of~\cite{Padovani:2014uq} are of further interest, as the authors point out that there is a significant correlation between several of the IceCube neutrino events and BL Lacs (3 events correlate with BL Lacs of known redshift, 4 events correlate with BL Lacs of unknown redshift).  We find these results very interesting.  When normalizing the BL Lac signal to the total IceCube neutrino flux we find that the number of neutrinos correlated with BL Lacs located closer than $z\sim0.2$ exceeds expectation by an order of magnitude.  A possible explanation of this discrepancy may be that the redshift distribution of BL Lacs capable of generating IceCube energy neutrinos is radically different from the redshift distribution of all BL Lacs due to some unknown mechanism.  However, our model suggests the mean free path of high energy neutrinos may not extend to high redshift due to absorption, potentially biasing the sources correlated with IceCube neutrino events to low redshift.  Beginning from this perspective, we define the BL Lac Source Correlation regime by supposing that absorption of high energy neutrinos on the C$\nu$B reconciles the over-abundance of correlated neutrino events at low {redshift}.  The mean free path of high energy neutrinos is shortened by scattering with secluded neutrinos in the C$\nu$B such that the correlation of 3 IceCube neutrinos with BL Lac sources at a redshift less than $z = 0.212$ (the largest redshift source correlated with an IceCube neutrino event~\cite{Padovani:2014uq}) is consistent with the total flux neutrinos observed by IceCube originating from all BL Lacs in the observable universe.  We also vary the expected number of background events in the IceCube 3 year data set by $1\sigma$~\cite{Aartsen:2014uq} to establish an upper and lower limit on the possible number of neutrinos originating form BL Lacs beyond a redshift of $z=.212$.  We show the results of these bounds in Figures~\ref{fig:gold} and~\ref{fig:projections}, {which correspond for $t$-channel scattering to an upper bound on the redshift horizon for high energy neutrinos of $z = 1.0\,(0.92)$ for the contact (continuum) limit and a lower bound on the redshift horizon of $z = 0.42\,(0.38)$ for the contact (continuum) limit.}}
 
\section{Discussion of Results}
\label{sec:discuss}

In Figs.~\ref{fig:gold} and~\ref{fig:projections} we summarize our main results. In each panel of Fig.~\ref{fig:gold} we fix the mediator mass and keep $g_{X} = g_{\nu}$ throughout. We include the thermal relic constraint, and exclude values of $g_{X}$ too small to yield a sufficiently large annihilation cross section (black curve). Strictly speaking this region of parameter space is only excluded in the minimal model here, since one could easily add new interactions to $X$ to open new annihilation channels. Next, we include DM self-interactions at the level of $\sigma_{XX}/m_{X} = 1-10~{\rm cm}^{2}{\rm g}^{-1}$ at dwarf scales where the DM dispersion is small, $v = 10~{\rm km}/{\rm s}$ (shaded blue). Self-Interactions at this level has been argued to provide a potential resolution of both the ``too-big-to-fail'' problem and the ``cusp-versus-core'' problems.  

In addition one may also wish to address the missing satellites problem. To this end in Fig.~\ref{fig:gold} we display curves of constant DM protohalo masses (green dashed curves). The observed lack of satellites in the Milky Way may be indicative of cutoffs around $10^{9-10}~M_{\odot}$. It is known however that DM substructure cannot be arbitrarily large. Lyman-$\alpha$ constraints are among the most restrictive and require $M_{halo} \lesssim 5\times 10^{10}~M_{\odot}$ (solid green)~\cite{SommerLarsen:1999jx,Tittley:1999jt}.  



{Significantly, the original region for the viable solution of all dark matter structure problems via hidden sector interactions with neutrinos~\cite{Aarssen:2012fx} persists within the larger parameter space allowed by our model.  Because the authors of~\cite{Aarssen:2012fx} explicitly considered only symmetric thermal relic dark matter which self-interacts in the classical scattering limit, their solution exists in the upper right hand portion of the plots in Figure~\ref{fig:gold}.  This subset of the allowed parameter space is almost entirely within the MFP$<50\,\rm Mpc$ regime for the IceCube experiment.  As such, the parameter space allowed by~\cite{Aarssen:2012fx} results in vigorous absorption of high energy neutrinos by the C$\nu$B, which can be readily probed by the IceCube detector.}  

To summarize the allowed parameter space we have projected the results of the calculations done for Fig.~\ref{fig:gold} onto the $m_\phi$ and $m_X$ planes in Fig.~\ref{fig:projections}.  These projections show that there is a very large volume of the parameter space which resolve the issues associated with dark matter structure while also producing absorption of high energy neutrinos by the C$\nu$B.  Most importantly, the right hand panel of  Fig.~\ref{fig:projections} shows that the bulk of the regime where IceCube might observe either isotropic or local high energy neutrino sources is also a viable solution for all of the dark matter structure problems.  This means that this model predicts, at the very minimum, a horizon in redshift beyond which high energy neutrinos cannot propagate to the IceCube detector without scattering and becoming undetectable.  Most of the parameter space produces an observable feature in the IceCube signal, specifically that the redshift distribution of high energy neutrino sources does not match the redshift distribution of observed events in the IceCube detector.  Furthermore, the possibility exists (dependent on the absolute neutrino mass scale) that absorption lines from resonant production of the mediator in $\nu - \nu$ collisions may be observable in the IceCube experiment. 

{Should the primary source of neutrinos in the IceCube detector ultimately prove to be BL Lacs and should the correlation found in~\cite{Padovani:2014uq} persist in the future, we} {could} {infer {data-preferred values} of the strength of the secluded sector self interactions.  From the BL Lac Source Correlation contour in Fig.~\ref{fig:projections}, we can make an estimate of the strength of the secluded sector interaction in the contact limit, $G_{\rm eff} = g_\nu^2/m_\phi^2 = 9 \pm 3 \times 10 ^{-5}\, {\rm MeV}^{-2}$, using $\sin\theta_s = 0.1$.}
\section{Conclusions}

We have studied the consequences of endowing DM with new interactions with neutrinos. This is a generic possibility in models where {secluded} neutrinos and DM are charged under a new $U(1)$ gauge symmetry. The mass mixing with active neutrinos arises from unspecified high-scale physics, which we effectively parameterize {as a} dimension-five operator connecting the two sectors. The new light gauge boson in this model simultaneously provides an annihilation channel for DM, yields late kinetic decoupling of DM and neutrinos, gives strong self-interactions at dwarf galaxy scales, {and modifies the mean free path of high-energy neutrinos.}  {Of further importance, DM is not self conjugate under such a scheme, allowing the DM relic abundance to be asymmetric.  This admits the existence of much more strongly coupled DM-DM and $\nu_s$-DM interactions, expanding the regime where such a model could simultaneously solve all of the dark matter structure problems} {and lead to novel effects at IceCube. On the neutrino side, the new interactions discussed here may have additional cosmological implications such as their impact on the CMB by delaying the onset of free-streaming~\cite{Bell:2005dr,Friedland:2007vv,Cyr-Racine:2013jua}.} {In addition, although present data has not yet reached thermal relic sensitivity~\cite{Abbasi:2011eq}, future IceCube data limiting DM annihilation into neutrinos will be a further test of this model, and can be relevant even in the case of asymmetric annihilation~\cite{Graesser:2011wi,Bell:2014xta}.}

{The neutrino mixing portal which connects the dark and SM sectors at low temperature has a range of beneficial effects.  Mixing angle suppression~\cite{Hannestad:2013ana,Dasgupta:2013zpn} precludes such models from interfering with BBN.  We have shown that late-time scattering via the dark mediator recouples the active and {secluded neutrino} populations, leading to a larger and hotter population of relic {secluded} neutrinos than previously thought~\cite{Dasgupta:2013zpn}.  This late time recoupling of neutrinos further increases the volume of the parameter space which may explain the missing satellites problem.}

{The most important feature present in our model is the $\nu$-$\nu$ scattering through the dark sector interaction which produces observable consequences in the high energy neutrino flux observed by the IceCube experiment.  IceCube data may be able to constrain or confirm $\nuDM$.  The presence of secluded neutrinos in the C$\nu$B modify the mean free path of high energy neutrinos {since they can} scatter on relic background of secluded neutrinos via mixing.  The parameter space which reconciles issues with dark matter structure lies almost entirely within the regime which might be tested as IceCube's exposure time increases.  The correlation of IceCube neutrino events with BL Lac sources at low redshift may be the first evidence of the absorption of high energy neutrinos through this mechanism. This represents a novel and unique opportunity to probe the dynamics of the {d}ark sector using the IceCube neutrino telescope.}


\vspace{.5cm}

\acknowledgements
{We would like to thank George Fuller and Kris Sigurdson for many helpful discussions and Bill Louis for comments on the manuscript. Results from this work have been presented at numerous conferences over the last year, including INFO13, MIAMI13, Neutrino 2014, SF14, Bright Ideas on Dark Matter at the University of Southern Denmark, and the NIAPP 2014 topical workshop. This work has been supported by the CP3-Origins centre which is partially funded by the Danish National Research Foundation, grant number DNRF90, the DOE Office of Science and the U.C. Office of the President in conjunction with the LDRD Program at LANL.  We would also like to thank the Munich Institute for Astro- and Particle Physics (MIAPP) and the Kavli Institute for Theoretical Physics at UCSB for their hospitality.}


\section*{Appendix: Kinetic Decoupling}
\vspace{1cm}

\subsection{Simplified Approach}
The matrix element for $ X \nu \rightarrow X \nu$ is (averaging over initial DM spins)
\be
\langle |\mathscr{M} |^{2}\rangle = \frac{4g_{X}^{2}g_{\nu}^{2}}{\left(t- m_{\phi}^{2}\right)^{2}} \left[ (s-m_{X}^{2})^{2} + (u- m_{X}^{2})^{2} + 2 m_{X}^{2} t\right] \nonumber
\ee
Note that using $s+t+u = 2m_{X}^{2}$ we have $u - m_{X}^{2} = -2m_{X}E_{\nu}-t$. 

The cross section is $d \sigma/d \cos \theta =\frac{ \langle |\mathscr{M} |^{2}\rangle }{32 \pi s}$. In the limit that $t =0$ and $s = m_{X}^{2} + 2m_{X}E_{\nu}$ we see that $\langle |\mathscr{M} |^{2}\rangle =  \frac{32 g_{X}^{2}g_{\nu}^{2}}{m_{\phi}^{4}}m_{X}^{2} E_{\nu}^{2}$ .

The temperature of kinetic decoupling can be found by solving $H = \Gamma_{mom} = n_{s} \sigma_{X\nu}\sqrt{\frac{3}{2}} \left(\frac{T_{s}}{m_{X}}\right)$ where at $t=0$ the cross section is $\sigma_{X\nu} =  \frac{g_{X}^{2}g_{\nu}^{2}E_{\nu}^{2}}{\pi m_{\phi}^{4}}$. Using $H = 1.66 \sqrt{g_{*}} T_{\gamma}^{2}/M_{Pl}$ and $n_{s} = \frac{3}{2} \frac{\zeta(3)}{\pi^{2}} N_{\nu} T_{s}^{3}$ and $\langle E_{\nu}^{2} \rangle = 12.9 T^{2}$. 
\be 
T_{KD} = \left(\frac{\pi^{3} 1.66 (2/3)^{3/2}}{12.9 \zeta(3) N_{\nu}}\right)^{1/4} \frac{m_{\phi} g_{*}^{1/8}}{\sqrt{g_{X}g_{\nu}}} \left(\frac{m_{X}}{M_{Pl}}\right)^{1/4}\left(\frac{T_{\gamma}}{T_{s}}\right)^{3/2}_{KD}. \nonumber
\ee

\subsection{Details}

The above approach is not fully accurate. A better method is offered by the following which takes into account Fermi-Dirac statistics and Pauli blocking. To do do we follow the method of~\cite{Gondolo:2012vh} and write the momentum-relaxation rate as 
\be \gamma(T) = \frac{g_{SM}}{6 m_{X}T} \int_{0}^{\infty} \frac{d^{3}p}{(2\pi)^{3}} f(p) \left(1-f(p)\right) 8 p^{4} \left. \frac{d \sigma}{dt}\right\vert_{t=0}
\ee
where $\frac{d\sigma}{dt} = \frac{1}{64 \pi m_{X}^{2} p^{2}} \langle |\mathscr{M} |^{2}\rangle$. 
\be 
 \left. \frac{d \sigma}{dt}\right\vert_{t=0} = \frac{g_{X}^{2}g_{\nu}^{2}}{2\pi m_{\phi}^{4}}
\ee
Using $\int_{0}^{\infty} p^{6} f(p) \left(1-f(p)\right)dp = \frac{31 \pi^{6}}{42} T^{7}$.  Using $g_{SM} =2 N_{\nu}$ we find 
\be 
\gamma(T) = \frac{31 N_{\nu}\pi^{3}}{63} \frac{g_{X}^{2}g_{\nu}^{2}}{m_{X}m_{\phi}^{4}} T_{\nu}^{6}
\ee
Equating this to the Hubble rate and solving for $T$ yields
\bea 
T_{KD} &=& \left(\frac{1.66 \cdot 63}{31 \pi^{3} N_{\nu}}\right)^{1/4} \frac{m_{\phi} g_{*}^{1/8}}{\sqrt{g_{X}g_{\nu}}} \left(\frac{m_{X}}{M_{Pl}}\right)^{1/4}\left(\frac{T_{\gamma}}{T_{s}}\right)^{3/2}_{KD}. \nonumber \\
&=& \frac{0.067~{\rm keV}}{N_{\nu}^{1/4}(g_{X}g_{\nu})^{1/2}} \left(\frac{T_{\gamma}}{T_{s}}\right)^{3/2} \left(\frac{m_{X}}{{\rm TeV}}\right)^{1/4} \left(\frac{m_{\phi}}{1~{\rm MeV}}\right) \nonumber,
\eea
which is in agreement with~\cite{Aarssen:2012fx}. 


\bibliography{asym_references.bib}

\end{document}